
\input harvmac

\def\inbar{\,\vrule height1.5ex width.4pt depth0pt}
\def\IB{\relax{\rm I\kern-.18em B}}
\def\IC{\relax\hbox{$\inbar\kern-.3em{\rm C}$}}
\def\ID{\relax{\rm I\kern-.18em D}}
\def\IE{\relax{\rm I\kern-.18em E}}
\def\IF{\relax{\rm I\kern-.18em F}}
\def\IG{\relax\hbox{$\inbar\kern-.3em{\rm G}$}}
\def\IH{\relax{\rm I\kern-.18em H}}
\def\II{\relax{\rm I\kern-.18em I}}
\def\IK{\relax{\rm I\kern-.18em K}}
\def\IL{\relax{\rm I\kern-.18em L}}
\def\IM{\relax{\rm I\kern-.18em M}}
\def\IN{\relax{\rm I\kern-.18em N}}
\def\IO{\relax\hbox{$\inbar\kern-.3em{\rm O}$}}
\def\IP{\relax{\rm I\kern-.18em P}}
\def\IQ{\relax\hbox{$\inbar\kern-.3em{\rm Q}$}}
\def\IR{\relax{\rm I\kern-.18em R}}
\font\cmss=cmss10 \font\cmsss=cmss10 at 7pt
\def\IZ{\relax\ifmmode\lrefhchoice
{\hbox{\cmss Z\kern-.4em Z}}{\hbox{\cmss Z\kern-.4em Z}}
{\lower.9pt\hbox{\cmsss Z\kern-.4em Z}}
{\lower1.2pt\hbox{\cmsss Z\kern-.4em Z}}\else{\cmss Z\kern-.4em Z}\fi}
\def\IGa{\relax\hbox{${\rm I}\kern-.18em\Gamma$}}
\def\IPi{\relax\hbox{${\rm I}\kern-.18em\Pi$}}
\def\ITh{\relax\hbox{$\inbar\kern-.3em\Theta$}}
\def\IOm{\relax\hbox{$\inbar\kern-3.00pt\Omega$}}

\newdimen\xraise\newcount\nraise
\def\xpoint{\hbox{\vrule height .45pt width .45pt}}
\def\udiag#1{\vcenter{\hbox{\hskip.05pt\nraise=0\xraise=0pt
\loop\ifnum\nraise<#1\hskip-.05pt\raise\xraise\xpoint
\advance\nraise by 1\advance\xraise by .4pt\repeat}}}
\def\ddiag#1{\vcenter{\hbox{\hskip.05pt\nraise=0\xraise=0pt
\loop\ifnum\nraise<#1\hskip-.05pt\raise\xraise\xpoint
\advance\nraise by 1\advance\xraise by -.4pt\repeat}}}
\def\bdiamond#1#2#3#4{\raise1pt\hbox{$\scriptstyle#2$}
\,\vcenter{\vbox{\baselineskip12pt
\lineskip1pt\lineskiplimit0pt\hbox{\hskip10pt$\scriptstyle#3$}
\hbox{$\udiag{30}\ddiag{30}$}\vskip-1pt\hbox{$\ddiag{30}\udiag{30}$}
\hbox{\hskip10pt$\scriptstyle#1$}}}\,\raise1pt\hbox{$\scriptstyle#4$}}

\def\inbar{\,\vrule height1.5ex width.4pt depth0pt}
\def\IB{\relax{\rm I\kern-.18em B}}
\def\IC{\relax\hbox{$\inbar\kern-.3em{\rm C}$}}
\def\IP{\relax{\rm I\kern-.18em P}}
\def\IR{\relax{\rm I\kern-.18em R}}
\def\inbar{\,\vrule height1.5ex width.4pt depth0pt}
\def\IB{\relax{\rm I\kern-.18em B}}
\def\IC{\relax\hbox{$\inbar\kern-.3em{\rm C}$}}
\def\ID{\relax{\rm I\kern-.18em D}}
\def\IE{\relax{\rm I\kern-.18em E}}
\def\IF{\relax{\rm I\kern-.18em F}}
\def\IG{\relax\hbox{$\inbar\kern-.3em{\rm G}$}}
\def\IH{\relax{\rm I\kern-.18em H}}
\def\II{\relax{\rm I\kern-.18em I}}
\def\IK{\relax{\rm I\kern-.18em K}}
\def\IL{\relax{\rm I\kern-.18em L}}
\def\IM{\relax{\rm I\kern-.18em M}}
\def\IN{\relax{\rm I\kern-.18em N}}
\def\IO{\relax\hbox{$\inbar\kern-.3em{\rm O}$}}
\def\IP{\relax{\rm I\kern-.18em P}}
\def\IQ{\relax\hbox{$\inbar\kern-.3em{\rm Q}$}}
\def\IR{\relax{\rm I\kern-.18em R}}
\font\cmss=cmss10 \font\cmsss=cmss10 at 7pt
\def\IZ{\relax\ifmmode\mathchoice
{\hbox{\cmss Z\kern-.4em Z}}{\hbox{\cmss Z\kern-.4em Z}}
{\lower.9pt\hbox{\cmsss Z\kern-.4em Z}}
{\lower1.2pt\hbox{\cmsss Z\kern-.4em Z}}\else{\cmss Z\kern-.4em Z}\fi}
\def\IGa{\relax\hbox{${\rm I}\kern-.18em\Gamma$}}
\def\IPi{\relax\hbox{${\rm I}\kern-.18em\Pi$}}
\def\ITh{\relax\hbox{$\inbar\kern-.3em\Theta$}}
\def\IOm{\relax\hbox{$\inbar\kern-3.00pt\Omega$}}

\def\Tr {\rm Tr}

\lref\alf{J. Alfaro, ``The large $N$ limit of the two hermitean
matrix model by the hidden BRST method'', CERN preprint
CERN-TH-6531-92, hep-th/9207097.}
\lref\bou{D.V. Boulatov, ``Infinite tension strings at $D>1$'',
N. Bohr Institute preprint NBI-HE-92-78, hep-th/9211064.}
\lref\stko{I.K. Kostov and M. Staudacher, Nucl. Phys. B384 (1992) 459.}
\lref\me{M.Staudacher, unpublished.}
\lref\dupko{B. Duplantier and I. Kostov, Nucl. Phys. B340 (1990) 491.}
\lref\bpiz{E. Br\'ezin, C. Itzykson, G. Parisi, and J.-B. Zuber,
Commun. math. Phys. 59 (1978) 35.}
\lref\grwi{D. Gross and E. Witten, Phys. Rev. D21 (1980) 446.}
\lref\mmdef{V. Kazakov, Phys. Lett. 150B (1985) 282;
F. David, Nucl. Phys. B 257 (1985) 45;
V. Kazakov, I. Kostov and A.A. Migdal, Phys. Lett. 157B (1985) 295;
J. Ambjorn, B. Durhuus, and J. Fr\"ohlich, Nucl. Phys. B 257 (1985) 433.}
\lref\nnprt{E. Br\'ezin and V. Kazakov, Phys. Lett. 236B (1990) 144;
M. Douglas and S. Shenker, Nucl. Phys. B335 (1990) 635;
D. Gross and A. Migdal, Phys. Rev. Lett. 64 (1990) 127.}
\lref\itzub{C. Itzykson and J.-B. Zuber, J. Math. Phys. 21 (1980) 411.}
\lref\mehta{M.L. Mehta, Comm. Math. Phys. 79 (1981) 327.}
\lref\migrev{For a review see: A.A. Migdal, Phys. Reports 102 (4) (1983).}
\lref\shat{S.L. Shatashvili,``Correlation functions in the
Itzykson-Zuber model'', Princeton-IAS preprint IASSNS-HEP-92-61,
hep-th/9209083.}
\lref\dan{D. Friedan, Commun. math. Phys. 78 (1981) 353.}
\lref\ade{I.K. Kostov, Nucl. Phys. B326 (1989) 583.}
\lref\katmm{V.A. Kazakov, Phys. Lett. 119A (1986) 140.}
\lref\kabou{D.V. Boulatov and V.A. Kazakov, Phys. Lett. 186B (1987) 379.}
\lref\ivamp{I.K. Kostov, Phys. Lett. 266B (1991) 317.}
\lref\goul{M. Goulian, Phys. Lett. 264B (1991) 292.}
\lref\mss{G. Moore, N. Seiberg and M. Staudacher,
Nucl. Phys. B 362 (1991) 665.}
\lref\kpz{V. Knizhnik, A. Polyakov and A. Zamolodchikov, Mod. Phys. Lett.
A3 (1988) 819.}
\lref\didk{F. David, Mod. Phys. Lett. A3 (1988) 1651;
J. Distler and H. Kawai, Nucl. Phys. B321 (1989) 509.}
\lref\mms{E. Martinec, G. Moore and N. Seiberg, Phys. Lett. B263 (1991) 190.}
\lref\cardy{J. Cardy, Nucl.Phys. B324 (1989) 581.}
\lref\mktda{M.Douglas, ``The two-matrix model'', 1990 Carg\'ese
workshop;
E.Martinec, Comm. Math. Phys. 138 (1991) 437;
T.Tada, Phys.Lett. B259 (1991) 442.}
\lref\daul{J.-M. Daul, V.A. Kazakov and I.K. Kostov, in preparation.}

\Title{RU-92-64}
{Combinatorial Solution of the Two-Matrix Model
}
\bigskip\centerline{Matthias Staudacher}

\centerline{Department of Physics and Astronomy}
\centerline{Rutgers University, Piscataway, NJ 08855-0849}

\vskip .3in

\baselineskip10pt{
We write down and solve a closed set of Schwinger-Dyson equations
for the two-matrix model in the large $N$ limit. Our elementary method
yields exact solutions for correlation functions involving
angular degrees of freedom whose calculation was impossible with
previously known techniques. The result sustains the hope that
more complicated matrix models important for lattice string theory and
QCD may also be solvable despite the problem of the angular
integrations. As an application of our method we briefly discuss
the calculation of wavefunctions with general matter boundary conditions
for the Ising model coupled to $2D$ quantum gravity.
Some novel insights into the relationship between lattice and continuum
boundary conditions are obtained.

\vskip 1cm
\leftline{Submitted for publication to {\it Physics Letters B}}

\Date{1/93}


\baselineskip=20pt plus 2pt minus 2pt

Matrix models in the large $N$ limit are widely believed to be
capable of giving qualitative and even quantitative insight into
non-abelian gauge theories.
The simplest such models provided early on very explicit solutions
to low-dimensional QCD \bpiz,\grwi. Over the past few years another physical
application of large $N$ matrix models has emerged: It has
become clear following \mmdef\
that hermitian models give a precise covariantly
regularized definition of perturbative bosonic string theory and $2D$ quantum
gravity. For the first time one was able to attempt to address
even non-perturbative issues in string theory \nnprt.
However the class of solved cases remains too small: The dimension of
the target spaces in which the world sheet (made up from the
matrix model Feynman diagrams) is embedded is too low.
In both the QCD as well as the string theory case
there is simply not enough room for the physically most interesting
excitations. It is well known that the the technical obstacle
preventing further progress consists in our inability to deal
with the relative ``angles'' between interacting matrices
in physically interesting multi-matrix models. This inability
to rid oneself of the angles would already be true in the simplest
interacting case --- the two-matrix model --- were it not for
the Itzykson-Zuber formula enabling one to actually
integrate them out \itzub. Unfortunately it appears impossible to
generalize this approach to any target space lattice containing
even one cycle. The question remains whether there does not exist
a general approach which does not break down once the angles
can no longer be eliminated.
In principle such a method is known: The loop (Schwinger-Dyson)
equations \migrev.
It is therefore of some interest to investigate whether the
two-matrix model can be understood in this language.
It will be demonstrated below that this is indeed the case.
After deriving
a set of closed loop equations using elementary combinatorial
reasoning we derive the algebraic (for polynomial potentials)
equation for the resolvent. This resolvent is called
``masterfield'' below since it encodes, via a Hilbert-transform,
the information about the spectral density of the matrices.
We then show how the general one trace correlators are expressed
as rational functions of the master field (further justifying
the use of language). We finally apply our formulae to
$2D$ quantum gravity and sketch how to extract some hitherto
unavailable information.

Consider then the two-matrix integral
\eqn\tmm{Z=\int{\cal D}A \ {\cal D}B \ e^{-N \Tr[V(A) + V(B) - c A B]}}
with $A$, $B$ hermitian in the large $N$ limit. $V(A)$ is a polynomial
potential which we will take for the most part\foot{ Everything works
equally well for more general polynomials and for the case
where $A$ and $B$ are controlled by non-equal potentials. Some
remarks on the first generalization follow below.}
to be $V(A)={1 \over 2} A^2 - {g \over 3} A^3$. The simplest ``observables''
in this model are the moments
\eqn\wn{W_n={1 \over N} \ \langle \ \Tr \ A^n \ \rangle}
where the expectation values $\langle \ldots \rangle$ are taken with respect
to the ensemble \tmm. These moments are the correlators calculable
with the standard technique of orthogonal polynomials \mehta.
This is not the
case for more complicated operator expectation values
\eqn\wnm{W^{(2k)}_{n_1,m_1,\ldots,n_k,m_k}={1 \over N} \ \langle
\ \Tr \ A^{n_1} B^{m_1} \ldots A^{n_k} B^{m_k} \ \rangle }
After changing variables to the eigenvalues of $A$ and $B$ a highly
nontrivial angular integration remains to be performed for each
correlator. Here we will find \wn,\wnm\ through a Schwinger-Dyson
technique\foot{Our method could be extended to the most general
connected correlators involving an arbitrary number of such
traces as well as to include $1 \over N^2$
corrections. This will not be done at present since our main point
is merely to demonstrate how to circumvent the ``angle'' problem.}.
It is useful to introduce the resolvent
\eqn\wp{W(P)={1 \over N} \ \langle \ \Tr \ {1 \over P-A} \ \rangle=
\sum_{n=0}^{\infty} \ ({1 \over P})^{n+1} \ W_n }
as well as the generating functions
\eqn\wkpq{\eqalign{W^{(2k)}(P_1,&Q_1,\ldots,P_k,Q_k) ={1 \over N}
\ \langle \ \Tr~ {1 \over P_1 - A}{1 \over Q_1 - B} \ldots
{1 \over P_k - A}{1 \over Q_k - B} \ \rangle \cr
&=\sum_{n_i,m_i=0}^{\infty} \ ({1 \over P_1})^{n_1+1} ({1 \over Q_1})^{m_1+1}
\ldots ({1 \over P_k})^{n_k+1} ({1 \over Q_k})^{m_k+1} \
W^{(2k)}_{n_1,m_1,\ldots,n_k,m_k} \cr } }
and the auxiliary functions
\eqn\wip{W_j(P)={1 \over N}\ \langle \ \Tr~ {1 \over P-A}~B^j~\rangle=
\sum_{n=0}^{\infty} \ ({1 \over P})^{n+1} \ W^{(2)}_{n,j}}
The strategy is now to write down recursion relations between the
coefficients \wn, \wnm\ and use them to find equations relating
the various generating functions. The recursion for $W_n$ is $(n \geq 1)$
\eqn\wnrc{W_n=g~W_{n+1} + c~W^{(2)}_{n-1,1} + \sum_{j=0}^{n-2}~ W_j~W_{n-2-j}}
Here $c$, $g$ are defined through eq.\tmm\ above.
Such recursions may be easily visualized and derived,
using elementary combinatorial arguments, by representing
the $W^{(2k)}$ by ``wheels'' with $2k$ alternating sequences of $n_1$ spokes
of color $A$, $n_2$ spokes of color $B$, and so on. Cutting out one spoke
and considering the possible new configurations gives the wanted
relations. The method is explained e.g. in \ade.
Eq.\wnrc\ implies, using \wp, the equation
\eqn\wpeq{(P-g~P^2)~W(P)=1 - g~(W_1 + P) + W(P)^2 + c~W_1(P)}
For $c=0$
(i.e. the one-matrix model) this is the well known basic loop equation
\migrev. In the two-matrix case the equation does not close
and we need to consider also two further recursions:
\eqn\wnerc{W^{(2)}_{n,1}=g~W^{(2)}_{n,2}+c~W_{n+1}}
and
\eqn\wnzrc{W^{(2)}_{n,1}=g~W^{(2)}_{n+1,1} + c~W^{(2)}_{n-1,2} +
\sum_{j=0}^{n-2}~W_{j}~W^{(2)}_{n-2-j,1} }
which are valid for $n \geq 0$ and $n \geq 1$, respectively.
Eqs.\wnerc,\wnzrc\ are summed up to give
\eqn\wpeeq{W_1(P)=g~W_2(P)+c~P~W(P)-c}
and
\eqn\wpzeq{(P-g P^2)~W_1(P)=(1-g P)~W_1-g W^{(2)}_{1,1}
+c W_2(P)+W(P)~W_1(P) }
Now eqs.\wpeq,\wpeeq,\wpzeq\ constitute a closed set of expressions for
the unknown functions $W(P),W_1(P),W_2(P)$ and we may deduce an
algebraic equation for the master field:
\eqn\mfld{\eqalign{W(P)^3 &+ \big[{c \over g} - 2(P-gP^2)\big] W(P)^2 +
\cr &+ \big[1-g(W_1+P) + {c^3 \over g} P - {c \over g} (P-g P^2)
+(P-gP^2)^2 \big] W(P) + \cr
&+ \big[(gW_1+gP-1)(P-gP^2) +
r_1 P + r_0
\big] = 0 \cr } }
It was found by different methods in \alf,\bou.
As in the case of the one matrix model master field equation
the remaining constants\foot{
One has
$r_1=-c+g W_1-g^2 W_2$ and $r_0=-g+{c \over g}-{c^3 \over g}-(1+c) W_1+
2 g W_2-g^2 W_3$ and
we have expressed $W^{(2)}_{1,1}$ through $W_1$,
$r_1$,$r_0$. }
$W_1$,$r_1$,$r_0$
are fixed by requiring that $W(P)$ possess
only one cut on the physical sheet.

Now let us extend these simple combinatorial arguments to derive some
novel results. The recursions \wnerc,\wnzrc\ are easily generalized
(for $n \geq 1$ and all $m \geq 0$) to
\eqn\wnmrc{W^{(2)}_{n,m}=g~W^{(2)}_{n+1,m} + c~W^{(2)}_{n-1,m+1} +
\sum_{j=0}^{n-2}~W_{j}~W^{(2)}_{n-2-j,m} }
Using eq.\wkpq\ one is led after some algebra to the equation
\eqn\wpqeq{W^{(2)}(P,Q)={ (1-g~P)~W(Q) - g~W_1(Q) - c~W(P) \over
P - g~P^2 - c~Q - W(P) } }
Thus, in view of eq.\wpeq,
the first nontrivial correlator involving ``angular'' degrees
of freedom is seen to be a simple rational function of the
master field $W(P)$.

Consider the next correlator $W^{(4)}$. The basic procedure
is identical to the one above, but the combinatorics gets
a bit more involved. The recursion is ($n_1 \geq 1$)
\eqn\frrc{\eqalign{W^{(4)}_{n_1,m_1,n_2,m_2}=& g~W^{(4)}_{n_1+1,m_1,n_2,m_2} +
c~W^{(4)}_{n_1-1,m_1+1,n_2,m_2} + \cr
&+ \sum_{j=0}^{n_1-2}~W_j~W^{(4)}_{n_1-2-j,m_1,n_2,m_2} +
\sum_{j=0}^{n_2-1}~W^{(2)}_{j,m_1}~W^{(2)}_{m_2,n_1+n_2-2-j} \cr } }
In order to be able to sum up this recursion and find an equation for
$W^{(4)}$ one needs a further relation (valid for all
$m_1,n_2,m_2 \geq 0$)
\eqn\mrrc{\eqalign{c~W^{(4)}_{1,m_1,n_2,m_2}=& W^{(2)}_{n_2,m_1+m_2+1} -
g~W^{(2)}_{n_2,m_1+m_2+2} - \cr
& -\sum_{j=0}^{m_1-1}~W_j~W^{(2)}_{n_2,m_1+m_2-1-j}
- \sum_{j=0}^{m_2-1}~W_j~W^{(2)}_{n_2,m_1+m_2-1-j} \cr } }
Then, using properties like
$W^{(4)}_{0,m_1,n_2,m_2}=W^{(2)}_{n_2,m_1+m_2}$, one derives after a
considerable amount of algebra
\eqn\wfreq{\eqalign{W^{(4)}&(P_1,Q_1,P_2,Q_2)=
{1 \over P_1 - gP_1^2 - cQ_1 - W(P_1)} \times \cr
& \times \bigg[ D_Q(Q_1,Q_2)\cdot \big[{g \over
c}\big(Q-gQ^2-W(Q_1)-W(Q_2)
\big)-1
+gP_1\big]~W^{(2)}(P_2,Q) +\cr
&~~~~~~ + \big(c-W^{(2)}(P_2,Q_1)\big)~D_P(P_1,P_2)\cdot W^{(2)}(Q_2,P) +
{g^2 \over c}~W(P_2) \bigg] \cr } }
where we have made use of the ``combinatorial derivatives''
\eqn\der{D_P(P_1,P_2)\cdot f(P) \equiv {f(P_2)-f(P_1) \over P_2 - P_1}}
It is now straightforward if somewhat tedious to generalize the
recursions \frrc,~\mrrc\ and thus compute all remaining
($k \geq 3$) one trace correlators $W^{(2k)}$:
\eqn\wkeq{\eqalign{W^{(2k)}(P_1,Q_1,\ldots,P_k,Q_k)=&
{1 \over P_1 - gP_1^2 - cQ_1 - W(P_1)} \times \cr
{}~~~~\times \Bigg[ D_Q(Q_1,Q_k)\cdot [{g \over
c}\big(Q-&gQ^2-W(Q_1)-W(Q_k)
\big)-1
+gP_1] \cdot \cr
&~~~~~~~~~~~~~~~~~ \cdot W^{(2k-2)}(P_2,Q_2,\ldots,P_k,Q) - \cr
- {g^2 \over c} D_P(P_2,P_k)\cdot W&^{(2k-4)}(P,Q_2,\ldots,Q_{k-1}) +\cr
{}~~~~+{g \over c} \sum_{l=2}^{k-1} (D_Q(Q_1,Q_l) \cdot
&W^{(2l-2)}(P_2,\ldots,P_l,Q) ) \cdot  \cr
&~~~~\cdot (D_Q(Q_l,Q_k) \cdot W^{(2k-2l)}(P_{l+1},\ldots,P_k,Q) ) + \cr
{}~~~+ cD_P(P_1,P_2) \cdot W&^{(2k-2)}(P,Q_2,\ldots,Q_k) - \cr
{}~~~~-\sum_{l=2}^{k}W^{(2l-2)}(P_l,Q_1&,\ldots,Q_{l-1})
D_P(P_1,P_l) \cdot W^{(2k+2-2l)}(P,Q_l,\ldots,Q_k) \Bigg] \cr  } }
We have thus succeeded in recursively expressing all higher
one trace correlators as rational functions of the master field
$W(P)$ alone. Note that the obtained formulae hide the fact that
the $W^{(2k)}$ possess a cyclic symmetry with respect to the
variables $P_i$,$Q_i$.

We now briefly sketch which modifications occur for a general polynomial
potential of order $m+1$:
$V(A)=-\sum_{r=2}^{m+1}{1 \over r} g_r A^r$ (here $g_2=-1$).
This generalization is of interest since any minimal $C<1$
conformal theory may be obtained by an appropriate\foot{
For non-unitary minimal models one actually needs unequal
potentials for the two matrices. We refrain from
giving formulae for that case.}
choice of $V(A)$ \mktda,\daul.
Studying the necessary recursions (their number increasing with the
order of the polynomial) one finds the masterfield equation
\eqn\mfldz{W(P)^{m+1} + h_m(P)~W(P)^m + \ldots + h_1(P)~W(P) + h_0(P) =
0}
The $h_i(P)$ are known polynomials in $P$, e.g.
$h_m(P)=-(c{g_m \over g_{m+1}} + mV'(P))$.
It is interesting to observe that the loop functions $W(P)$ are
{\it algebraic}, just as in the case of the $O(n)$ model at rational points
\stko. One again finds that the higher one-trace correlators are
rational functions of the master field; e.g. $W^{(2)}$ reads
\eqn\rat{W^{(2)}(P,Q)={-\sum_{r=1}^{m}~\sum_{s=1}^{r}~g_{r+1}~P^{r-s}~
W_{s-1}(Q) - c~W(P) \over
V'(P) - c~Q - W(P) } }
The functions $W_i(P)$ are related to each other and $W(P)$
through expressions similar to \wpeq; these relations are derived
using the same methods
as above. Alternatively one can find them by investigating \rat\
for $Q \rightarrow \infty$ and using \wkpq.

Let us complement this analysis with a brief application of our
results to continuum $2D$ quantum gravity.
It is well known that the two-matrix theory eq.\tmm\ describes at its
critical point the $C={1 \over 2}$ conformal theory coupled to
$2D$ gravity \katmm,\kabou. Indeed, analyzing the master field equation
\mfld\ one reproduces the correct critical values\foot{
They are $c_{*}=0.1589\ldots$, $g_{*}=0.2004\ldots$, $P_{*}=3.6850\ldots$.}
for $c$,$g$,$P$ \kabou.
Scaling $g=g_{*}-a~\mu$ and $P=P_{*}+a~z$ where $a$ is the cutoff
and $\mu$ and $z$ are the continuum bulk and boundary cosmological
constants one reproduces the correct tree-level one-boundary
amplitude \ivamp,\goul,\mss\ (in the first reference this result was obtained
from a different formulation of the $C={1 \over 2}$ theory and in the
latter references from the same model, but in Laplace-transformed form):
\eqn\scw{W(P)~\sim~a^{{4 \over 3}}~w(z)~=~
a^{{4 \over 3}}~\big[(z+\sqrt{z^2-\mu})^{{4 \over 3}}
+(z-\sqrt{z^2-\mu})^{{4 \over 3}}\big]}
Our calculation of the general moments \wnm\ enables one to now study
a much richer set of matter boundary conditions. However, even though
eqs.\wpqeq,\wfreq,\wkeq\ are rather explicit rational
functions of the amplitude
$W(P)$ the scaling of these expressions is more subtle than one
might expect at first. In particular, the continuum limits of the
amplitudes $W^{(2k)}$ are not simply rational functions of the scaled
loop function $w(z)$. Instead one has to expand $W(P)$ to higher powers
in the cutoff $a$ than order ${{4 \over 3}}$; i.e.~one needs to keep
terms that were neglected in \scw. Here we will only present results
for $\mu=0$; one finds, scaling\foot{Note that one could also weigh
each boundary segment with its own boundary cosmological constant.}
$P_i=P_{*}+az$, $Q_i=P_{*}+az$
\eqn\scal{W^{2k}(P_1,Q_1,\ldots,P_k,Q_k)~\sim~(a z)^{{7/3-k}}}
It is instructive to compare this scaling law to the continuum
KPZ/DDK prediction for the insertion of $2k$ boundary operators
${\cal O}_i$ into a disc ($\varphi$ is the Liouville field,
$\gamma=\sqrt{{2 \over 3}}$ and $\alpha_i$ the dressing charge of
${\cal O}_i$) \kpz,\didk,\mms:
\eqn\ddk{\bigg\langle
\prod_{i=1}^{2k}~\big(\oint {\cal O}_i e^{\alpha_i \varphi}\big)
\bigg\rangle~\sim~z^{{7 \over 3}-{2 \over \gamma} \sum_{i=1}^{2k}
\alpha_i}}
Therefore one may deduce the charge of the continuum boundary operator
${\cal O}$ producing the behavior \scal\ to be
${2 \alpha \over \gamma}={1 \over 2}$. One then finds, using
${2 \alpha \over \gamma}={1 \over 6} (7-\sqrt{1+48\Delta})$
\kpz,\didk,\mms, the flat
space boundary scaling dimension $\Delta$ of ${\cal O}$:
\eqn\del{\Delta({\cal O})~=~{5 \over 16}}
This is not a dimension occuring
in the minimal Kac table of the Ising model. It appears however
in the extended Kac table, suggesting that it does correspond
to a nonlocal boundary operator in a non-minimal extension of the
$C={1 \over 2}$ conformal theory. The result might be
surprising at first: An operator $\langle \Tr A^{n} B^{m}\rangle$
would naively correspond to a boundary with a sequence of up
spins followed by a sequence of down spins. It is well known
that such a spin configuration corresponds to the insertion of
two spin operators $\sigma$ into the boundary \cardy. But
the boundary dimension of $\sigma$ is $\Delta(\sigma)={1 \over 2}$
as opposed to \del. We believe that the key to resolving this puzzle
lies in considering more carefully the original random surface
interpretation of the theory \tmm. For closed surfaces we can
imagine the Ising spins to sit either on the vertices of the
$\varphi^3$-lattice or alternatively on the vertices of the
triangulation dual to the $\varphi^3$-lattice:
Kramers-Wannier duality holds also for random lattices.
Once we include
boundaries it is easily seen that insertion of the operator
$\Tr (A^n+B^n)$ into \tmm\ corresponds to free boundary conditions
for the spins on the triangulation. An operator like
$\langle \Tr A^{n} B^{m}\rangle$, on the other hand, does not correspond
to a simple local boundary condition for the Ising spins on the
triangulation. In short, we would like to argue that
Kramers-Wannier duality is destroyed once we consider open surfaces
and that we should imagine the spins to sit on the lattice dual
to the $\varphi^3$ lattice. Further evidence for this interpretation may
be obtained from the loop gas approach \ade\ which
enables one to consider boundaries with $2k$ alternating
segments of up and down spins (situated at the vertices of the
triangular lattice) separated by domain walls. A simple
analysis \me\ of the disc endowed with these boundary conditions
along the lines of \dupko\ gives the behavior
$\sim z^{{7 \over 3}-{2 \over 3} k}$.
This agrees (using $\Delta(\sigma)={1 \over 2}$) with the
continuum prediction
\eqn\spin{\bigg\langle
\big(\oint \sigma e^{\alpha_i \varphi}\big)^{2k}
\bigg\rangle~\sim~z^{{7 \over 3}-{2 \over 3} k}}
Obviously further results (e.g. for $\mu \neq 0$ and for the
higher critical points)
could be derived from the above formulae.

We have thus demonstrated that the master field of the
two-matrix model with polynomial potential is an algebraic
function and that the one-trace correlators are rational
functions of the master field. It is not easy to imagine
how to obtain the latter result using the standard method of
orthogonal polynomials: Some rather complicated angular
integrations have to be performed (but see \shat).
The most interesting question is of course whether the study of the loop
equations of more complicated matrix models will lead to
new solutions. It would be of some interest to understand for
which target space lattices the master field, be it for
hermitian or unitary matrices\foot{
We remind the reader that for sufficiently simple link structures
unitary models possess an algebraic spectral distribution
as well \grwi,\dan.
Of course we know
that already a model consisting of four linearly interacting
hermitean matrices
can not possibly have an algebraic master field: Corresponding
to a $C=1$ lattice theory (the 4-state Potts model),
it will certainly have a logarithmic singularity in its eigenvalue density.},
remains algebraic. There are rather ``small'' matrix models like
hermitean five matrix models and unitary Eguchi-Kawai type models \migrev\
whose solution would presumably lead to important new physical
insights: The former would contain a $C > 1$ critical point in bosonic
string theory and the latter would give an analytic expression for
strongly coupled QCD. Is it truly impossible to find the master field
for these models?

\bigbreak\bigskip\bigskip\centerline{{\bf Acknowledgements}}\nobreak
We would like to thank M.Douglas for discussions. This work was supported
in part by DOE grant DE-FG05-90ER40559.

\listrefs


\end